\documentclass[aps,prb,twocolumn,superscriptaddress,showpacs]{revtex4}
\usepackage{graphics}
\usepackage{graphicx}
\usepackage[english]{babel}
\usepackage{color}

\begin{document}

\title{Vortex matter freezing in Bi$_{2}$Sr$_{2}$CaCu$_{2}$O$_{8}$
samples with a very dense distribution of columnar defects}


\author{N. R. Cejas Bolecek}
\affiliation{Low Temperature Lab, Centro At\'{o}mico Bariloche \&
Instituto Balseiro, Bariloche, Argentina}
\author{ A. B. Kolton}
\affiliation{Solid State Theory Group, Centro At\'{o}mico
Bariloche \& Instituto Balseiro, Bariloche, Argentina}
\author{M. Konczykowski}
\affiliation{Laboratoire des Solides Irradiées, CNRS UMR 7642 \&
CEA-DSM-IRAMIS, Ecole Polytechnique, Palaiseau, France}
\author{H. Pastoriza}
\affiliation{Low Temperature Lab, Centro At\'{o}mico Bariloche \&
Instituto Balseiro, Bariloche, Argentina}
\author{ D. Dom\'{\i}nguez}
\affiliation{Solid State Theory Group, Centro At\'{o}mico
Bariloche \& Instituto Balseiro, Bariloche, Argentina}
\author{Y. Fasano}
\affiliation{Low Temperature Lab, Centro At\'{o}mico Bariloche \&
Instituto Balseiro, Bariloche, Argentina}


\date{\today}

\begin{abstract}
We show that the dynamical freezing of vortex structures nucleated
at diluted densities in Bi$_{2}$Sr$_{2}$CaCu$_{2}$O$_{8}$ samples
with a dense distribution of columnar defects, $B \sim 10^{-2}
B_{\Phi}$ with $B_{\Phi}=5$\,kG, results in configurations with
liquid-like correlations. We propose a freezing model considering a
relaxation dynamics dominated by double-kink excitations driven by
the local stresses obtained directly from experimental images. With
this model we estimate the relaxation barrier and the freezing
temperature. We argue that the low-field frozen vortex structures
nucleated in a dense distribution of columnar defects thus
correspond to an out-of-equilibrium non-entangled liquid with
strongly reduced mobility rather than to a snapshot of a metastable
state with divergent activation barriers as for instance expected
for the Bose-glass phase at equilibrium.

\end{abstract}

 \keywords{}

 \maketitle

\section{Introduction}

Vortices nucleated in high-temperature superconducting
samples~\cite{Blatter} are paradigmatic systems to study the phases
frozen in substrates with strong disorder since the relevant
energies are easily tuned by changing temperature and magnetic
field. Direct imaging techniques~\cite{Fasano2005,Petrovic2009}
allow a quantitative analysis of the impact of disorder on the
otherwise perfect equilibrium Abrikosov lattice. Magnetic decoration
studies~\cite{Fasano1999,Fasano2003} of quenched vortex structures
unveiling a large number of vortices has become a promising avenue
to perform these
studies.~\cite{Demirdis2011,Demirdis2012,Demirdis2013,Yang2012,vanderbeek2012,vanderbeek2013}
 This technique provides a two-dimensional top view at the sample
surface of the three-dimensional  vortex lattice frozen at $T_{\rm
freez}$. In order to move forward on the quantification of the
impact of disorder introduced by defects in the vortex structure,
better understanding or modelling of the vortex dynamics during the
cooling process is necessary. In this work we address this issue by
using vortex-defect interaction force magnitudes obtained from
magnetic decoration results as input to a theoretical freezing model
considering strongly localized vortices.

In field-cooling magnetic decorations, a high-temperature vortex
state is driven into a frozen configuration at $T_{\rm freez}$, an
intermediate temperature between the initial state and the lower
temperatures at which experiments are
performed.~\cite{Fasano1999,Fasano2003} Having a general
quantitative understanding of this freezing process is difficult due
to the non-equilibrium and non-stationary nature of the
three-dimensional thermally-activated vortex dynamics over pinning
barriers. In order to undergo this study is then desirable to find a
convenient experimental situation in which the vortex dynamics
modelling could be simplified. Therefore we have chosen to study the
case of a  dense distribution of strong pinning centers such as
columnar defects (CD), known to prevail over any other type of
cristalline
disorder.~\cite{Civale1990,Konczykowski1991,Keeshabilitation}
Vortices are then expected to become individually localized at CD
where their thermally-activated motion can be modelled in terms of
simple excitations, which drive the system into the putative
equilibrium Bose-glass phase.~\cite{Nelson1993} Although
experimental evidence for the Bose-glass dynamics was reported, the
equilibration time  and whether the frozen structures observed by
magnetic decoration can reveal aspects of this phase for a dense
distribution of CD remain as important open questions.

In this work we study via magnetic decoration the structural
properties of  vortex matter nucleated in
Bi$_{2}$Sr$_{2}$CaCu$_{2}$O$_{8}$
 samples with a large ratio of CD to vortices, $n_{\rm CD}/n_{\rm
v}=B_{\rm \Phi}/B$, with $B_{\rm \Phi}=n_{\rm CD}\Phi_{0}=5$\,kG the
matching field. Although transport and magnetic relaxation
experiments were performed for the same
system~\cite{vanderbeek1995,vanderbeek1995b,Soret2000}, the
structural properties were previously studied only for smaller doses
of CD. A complete destruction of the positional and orientational
order of the vortex structure is
reported.~\cite{Leghissa1993,Dai1994,Menghini2003,Banerjee2003,Menghini2004,Banerjee2004,Fasano2004,Menghini2005}
However, for very low CD densities of tens of Gauss, the short-range
order of the vortex structure is recovered \cite{Dai1994} and a
polycristalline structure is observed in magnetic decoration
snapshots
~\cite{Menghini2003,Banerjee2003,Menghini2004,Banerjee2004,Fasano2004,Menghini2005}
indicating inter-vortex repulsion remains important even in the
presence of such strong pins.  As mentioned,  the vortex
configurations imaged by field-cooling decorations correspond to the
state frozen at $T_{\rm freez}$ at which vortex mobility is strongly
reduced by the effect of pinning.~\cite{Pardo1997,Fasano2005} Here
we show that the quenched structures can be well described with a
freezing model that considers that the relaxation dynamics of
vortices in a dense CD potential is mainly dominated by double-kink
excitations. We estimate the value of the relaxation energy barriers
from data of the maximum vortex-defect interaction force obtained
from magnetic decoration experiments. This allows us to argue that
the imaged vortex structures correspond to a experimental-time
resolution limited frozen non-entangled liquid rather than to a
snapshot of the Bose-glass phase characterized by metastable states
separated by divergent barriers.

\section{Experimental}

The single-crystal of optimally-doped
Bi$_{2}$Sr$_{2}$CaCu$_{2}$O$_{8}$ studied here was grown by the
travelling-floating-zone method \cite{Li1994} and irradiated by 6
GeV Pb-ions at GANIL. The irradiation dose was chosen in order to
obtain a density of $2.42\,\times\,10^{10}$ CD per square
centimeter corresponding to a matching field of $B_{\rm
\Phi}=5$\,kG. Every single ion impact creates an amorphous
columnar track with a radius $r_{\rm r}\sim 3.5$\,nm, roughly
parallel to the c-axis through the entire sample thickness.  This
gives a bare-pinning energy per unit length per CD at zero
temperature of $U_{\rm 0} = \varepsilon_{\rm 0}\ln{(r_{\rm
r}/\sqrt{2}\xi(0))} \sim 6 \varepsilon_{\rm 0}(0) \sim 4 \times
10^{-6}$\,erg/cm.

The studied crystal was characterized by means of differential
magneto-optics,~\cite{Dorosinskii1992,Demirdis2011} local
Hall-probe magnetometry,~\cite{Konczykowski1991b} and magnetic
decoration~\cite{Fasano2008} techniques. The sample has a critical
temperature $T_{\rm c}=86.7$\,K, in-plane dimensions of $400\,
\times\, 200\,\mu$m$^{2}$, and thickness of some tens of microns.
Magneto-optical imaging shows rather homogeneous flux penetration
into the crystals and does not reveal any noticeable large-scale
surface or bulk defect.

\begin{figure}[bbb]
\includegraphics[width=\columnwidth,angle=0]{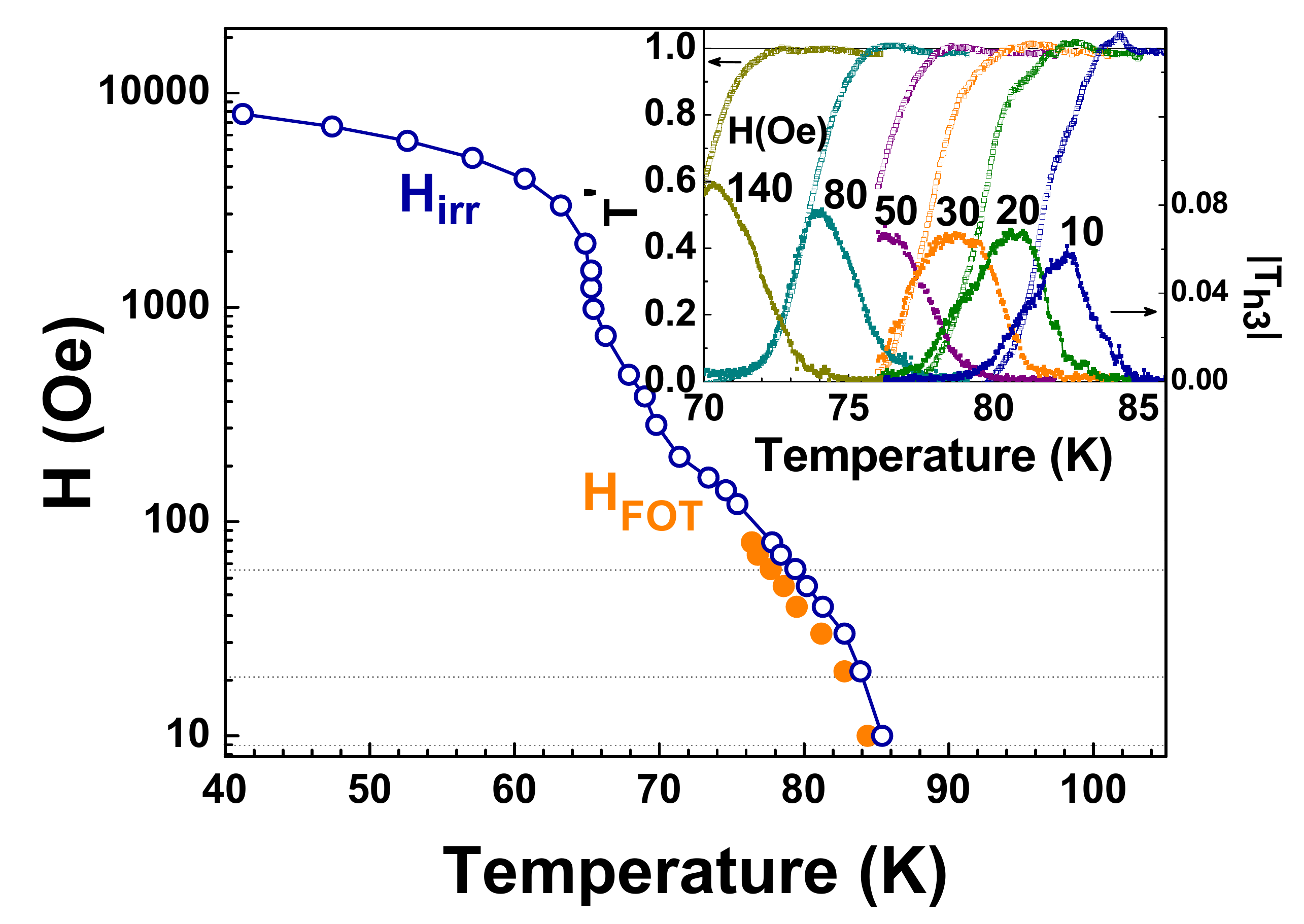}
\caption{Vortex phase diagram for Bi$_{2}$Sr$_{2}$CaCu$_{2}$O$_{8}$
with a dense distribution of CD ($B_{\rm \Phi}=5$\,kG).
Irreversibility, $H_{\rm irr}$, and first-order, $H_{\rm FOT}$,
transition lines obtained from transmittivity, $T'$, and modulus of
the third-harmonic response, $\mid T_{\rm h3} \mid$, as explained in
the text. Dotted lines indicate the field-cooling processes followed
during magnetic decorations. Insert: Temperature-dependence of $T'$
and $\mid T_{\rm h3} \mid$. \label{figure1}}
\end{figure}

The $H$\,-\,$T$  phase diagram of Bi$_{2}$Sr$_{2}$CaCu$_{2}$O$_{8}$
vortex matter with such a dense distribution of CD  was obtained by
means of local Hall-probe magnetometry up to $10^5$\,G, see
Fig.\,\ref{figure1}. These measurements were done using
microfabricated 2D-electron-gas Hall-magnetometers that locally
probe the sample stray field.~\cite{Konczykowski1991b} Magnetic
transmittivity measurements were performed by applying an ac
excitation field $H_{\rm ac}$ parallel to a dc field $H$. The Hall
data presented here were obtained with an excitation field of
$1.2$\,Oe rms and 11\,Hz. A digital-signal-processing lock-in
technique is used to simultaneously measure the in- and out-of-phase
components of the fundamental and the third-harmonic signals of the
Hall voltage. The fundamental signal was used to obtain the
thermodynamic first-order transition
line,~\cite{Pastoriza1994,Zeldov1995} $H_{\rm FOT}(T)$, from
transmittivity $T'$ measurements.~\cite{Konczykowski2006} The
third-harmonic signal $\mid T_{\rm h3} \mid$ yields information on
the onset of irreversible magnetic behavior at $H_{\rm
irr}(T)$.\cite{Konczykowski2006}

The structural properties of vortex matter nucleated on the same
crystal  were directly imaged by means of magnetic decoration
experiments.~ \cite{Fasano2008}  This study was limited to magnetic
fields below 80\,Oe since for this sample  the technique looses
single-vortex resolution at larger vortex densities. The sample was
field-cooled from $T>T_{\rm c}$ down to 4.2\,K in roughly 15\,min
and magnetic decorations were performed at this base temperature.
The structural properties, at the lengthscales of the lattice
parameter, correspond to those frozen at $T_{\rm freez}$. By using
quantitative information obtained from magnetic decoration images we
will estimate $T_{\rm freez}$ according to the freezing dynamics
model presented in Section\,IV.

\section{Experimental results}

\subsection{Vortex phase diagram}

Figure\,\ref{figure1} shows the vortex phase diagram of
Bi$_{2}$Sr$_{2}$CaCu$_{2}$O$_{8}$ vortex matter nucleated in samples
with a CD  density of $B_{\rm \Phi}=5$\,kG. The first-order, $H_{\rm
FOT}$, and irreversibility, $H_{\rm irr}$, lines are obtained from
measuring the sample magnetic response by means of ac Hall
magnetometry.~\cite{Konczykowski2006} The insert shows the
temperature evolution of normalized first and third-harmonic signals
in the low-field range.  The transmittivity $T'$ is obtained from
the in-phase component of the first-harmonic
signal,~\cite{Gilchrist1993} and is highly-sensitive to
discontinuities in the local induction, as for example the one
entailed at the $H_{\rm FOT}$
transition.~\cite{Pastoriza1994,Zeldov1995} The normalized modulus
of the third-harmonic signal,  $\mid T_{\rm h3}
\mid$,~\cite{Gilchrist1993} becomes non-negligible at the onset of
non-linear response arising from irreversible magnetic properties.

\begin{figure*}
\includegraphics[width=2\columnwidth]{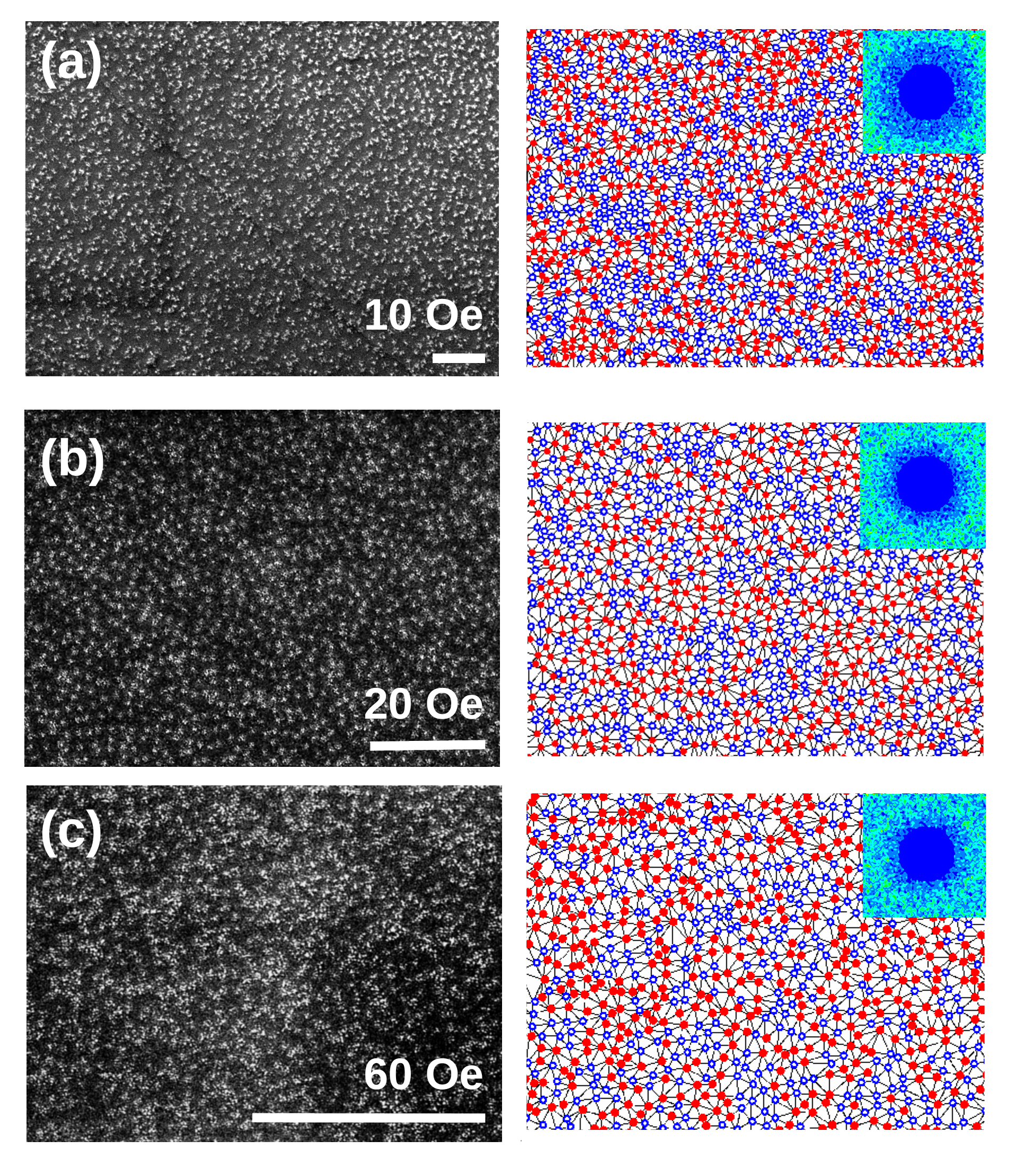}
\caption{Vortex structure in Bi$_2$Sr$_2$CaCu$_2$O$_8$ samples
with a CD density corresponding to $B_{\rm \Phi}=5$\,kG for
applied fields of (a) 10, (b) 20 and (c) 60\,Oe. Left panels:
Magnetic decoration images of the vortex structure taken at 4.2\,K
after field-cooling from the liquid vortex phase. The white bars
correspond to  10\,$\mu$m. Right panels: Delaunay triangulations
indicating first-neighbors
 and sixfold (blue) and non-sixfold (red) coordinated
vortices. The inserts show the Fourier transform of the vortex
positions.
 \label{figure2}}
\end{figure*}

The high-temperature $H_{\rm FOT}$ transition is detected in ac
transmittivity measurements as a frequency-independent so-called
paramagnetic peak that develops in $T'$ at the same $H$ as the jump
in local induction detected in dc hysteresis
loops.~\cite{Morozov1996,Konczykowski2006} The paramagnetic peak is
equivalently observed in $T'$ versus temperature curves, see
Fig.\,\ref{figure1}. For the studied sample the paramagnetic peak is
clearly observed in $T'$ curves up to 80\,Oe. The irreversibility
line is identified from the frequency-dependent onset of $\mid
T_{\rm h3} \mid$ on cooling.~ \cite{Dolz2014}  In the range $H <
B_{\rm \Phi}/6$, $T_{\rm irr}$ monotonically shifts towards lower
temperatures on increasing field, whereas for $B_{\rm \Phi}/6 < H
<B_{\rm \Phi}/3$ becomes almost field-independent.
 At larger fields, a monotonous increase of $H_{\rm
irr}$ with reducing field is again observed. This field-evolution of
the irreversibility line is common to
Bi$_{2}$Sr$_{2}$CaCu$_{2}$O$_{8}$ samples  with high-densities of CD
and has origin in the three different regimes for the occupation of
columnar defects with vortices discussed in
Refs.\,\onlinecite{vanderBeek2000,vanderbeek2001}.

Figure\,\ref{figure1} also  indicate with dotted lines the
$H$\,-\,$T$ paths followed during the field-cooling magnetic
decoration experiments. For all the imaged vortex structures the
system undergoes the melting transition at $T_{\rm FOT} \sim 0.98
T_{\rm irr}$. This shifting between both lines, although of lesser
intensity, was also reported in pristine samples.~\cite{Dolz2014}
\color{black}

\subsection{Structural properties of the frozen vortex matter}

\begin{figure}
\includegraphics[width=\columnwidth]{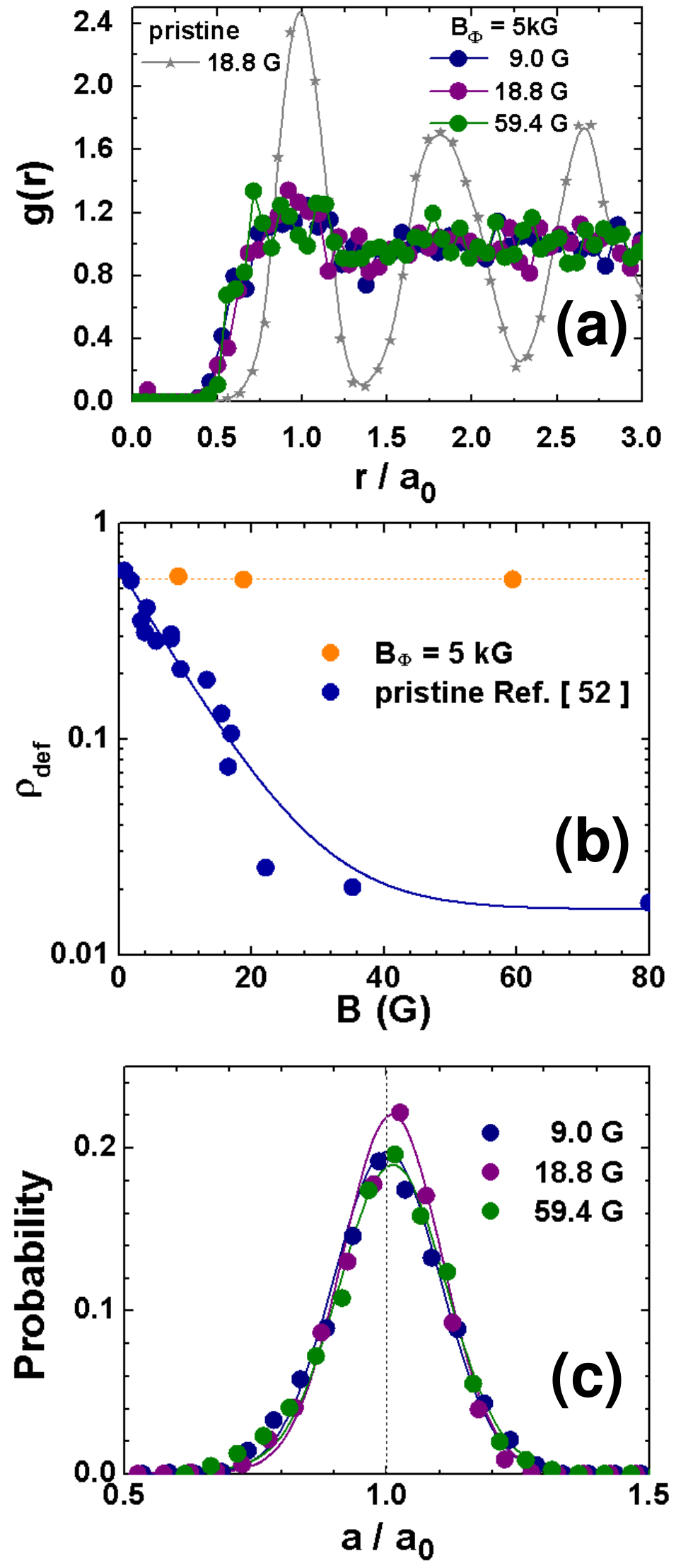}
\caption{Structural properties of
Bi$_{2}$Sr$_{2}$CaCu$_{2}$O$_{8}$ vortex matter nucleated in
samples with a dense distribution of CD corresponding to $B_{\rm
\Phi}=5$\,kG. (a) Pair-correlation function for the vortex
structures nucleated at several fields (circles). For comparison,
we show the results in a pristine sample at 18.8\,G. (b)
Field-evolution of the density of non-sixfold coordinated vortices
obtained from the images of Fig.\,\ref{figure2} as compared to the
case of vortex matter nucleated in pristine samples. Lines are
guides to the eye. (c) Distribution of first-neighbors distances,
$a$, normalized by the average lattice spacing $a_{0}$ (points).
The full lines are fits to the data with Gaussian functions.
 \label{figure3}}
\end{figure}

Figure\,\ref{figure2} shows snapshots of the vortex structure
obtained in field-cooling magnetic decoration experiments at applied
fields of 10, 20 and 60\,Oe. Due to the finite magnetization of
these samples with pinning enhanced by CD, the local induction
measured from the vortex density is smaller than the applied field,
$B=9$, 18.8, and 59.4\,G, respectively. In this field range vortices
are extremely diluted with respect to the random distribution of CD,
every vortex unit cell spanning a spatial region having between 80
(60\,Oe) and 500 (10\,Oe) defects. In this limit, one can expect
that the vortex structure presents similar topological order than in
the case of pristine samples. Strikingly, the observed vortex
structures are amorphous, irrespective of the vortex density within
the studied range. These results follow the same trend than magnetic
decoration data on samples with a less-dense distribution of CD
($B_{\rm \Phi}=3.5$\,kG) were a complete destruction of
translational order was reported. \cite{Leghissa1993}

The Delaunay triangulations of the right panels of
Fig.\,\ref{figure2}, indicating the first-neighbors for each vortex,
reveal that the structures have the non-sixfold coordination typical
of an amorphous. The ring-like patterns of the Fourier transform of
vortex positions shown in the insert confirm the lack of short-range
positional and orientational orders. Indeed, the pair correlation
functions $g(r)$ of Fig.\,\ref{figure3}(a) show only one
distinguishable peak at the first-neighbors distance, irrespective
of significantly increasing the inter-vortex interaction. This
contrasts with the $g(r)$ obtained for pristine samples that present
peaks up to several lattice spacings (see the curve with stars in
Fig.\,\ref{figure3} (a)). The sixfold coordinated vortices (blue)
form very small crystallites containing at best 10 vortices. The
density of these vortices is always below 40\,\% and does not vary
significantly on increasing field, see Fig.\,\ref{figure3} (b).
Quantitatively similar results are obtained for samples with a much
more diluted distribution of CD when $\sqrt{B/B_{\rm \Phi}} <
1.1$.~\cite{Menghini2004} Figure \ref{figure3} (b) also shows that
the density of vortices belonging to topological defects, $\rho_{\rm
def}$,  is significantly larger in samples with a dense distribution
of CD than in the case of vortex matter nucleated in pristine
Bi$_{2}$Sr$_{2}$CaCu$_{2}$O$_{8}$ samples.~\cite{Fasano2005}
However, since magnetic decoration images are snapshots of the
vortex structure, distinguishing between an amorphous glassy and a
liquid vortex phase is not possible with this technique.

Similar amorphous  structures were recently reported for the
strongly-pinned  vortex matter in pnictide
BaFe$_{2}($As$_{1-x}$P$_{x})_2$ and
Ba(Fe$_{1-x}$Co$_{x}$)$_{2}$As$_{2}$ samples,
~\cite{Demirdis2011,Demirdis2013} presenting significant
vortex-density fluctuations, more pronounced in the Co-doped system.
The disorder of the vortex structure was quantitatively ascribed to
the strong inhomogeneous disorder present in the samples.
\cite{Demirdis2011} On the contrary, in the
Bi$_{2}$Sr$_{2}$CaCu$_{2}$O$_{8}$ samples studied here, vortex
density fluctuations are not very strong, as observed in the
histograms of $a/a_{\rm 0}$ shown in Fig.\,\ref{figure3} (c). These
histograms are well fitted by a symmetric Gaussian distribution with
full-width at half-maximum ranging 23-20\,\%. The data are
normalized by the lattice parameter $a_{\rm 0}=1.075\sqrt{\Phi_{\rm
0}/B}$, with $B$ obtained from the vortex density measured in
magnetic decoration images.

In the field-cooling magnetic decoration experiments performed here,
vortex matter is nucleated in the high-temperature liquid phase and
vortices have a high mobility since the decoupled pancakes present a
low shear viscosity. On cooling, vortex mobility gets reduced at
$T_{\rm freez}$ by the effect of the CD pinning potential and the
vortex structure gets frozen, at lengthscales of the lattice
parameter $a_{0}$, in one of the many available metastable states.
On further cooling to lower temperatures, vortices accommodate in
order to profit from the pinning potential in all their length at
lengthscales of the order of coherence length, $\xi \ll a_{0}$, a
lengthscale that can not be resolved by means of magnetic
decoration. As a consequence, the structural properties of vortex
matter revealed by magnetic decorations at 4.2\,K correspond to
those of the structure frozen at $T_{\rm freez}$. Therefore this is
the temperature that has to be considered in order to evaluate
elastic and electromagnetic properties of vortex matter observed by
means of magnetic decoration. As usually considered in the
literature \cite{Fasano2005,Pardo1997}, it is reasonable to assume
$T_{\rm freez} \sim T_{\rm irr}$ since at this  temperature pinning
becomes dominant over the other energy scales. However, there can be
a shift between these two temperatures: The  model presented in
Section IV will allow us to discuss on the validity of this
assumption and its implications.

\subsection{Vortex-defect force distributions}

\begin{figure}[bbb]
\hspace{-1cm}
\includegraphics[width=\columnwidth]{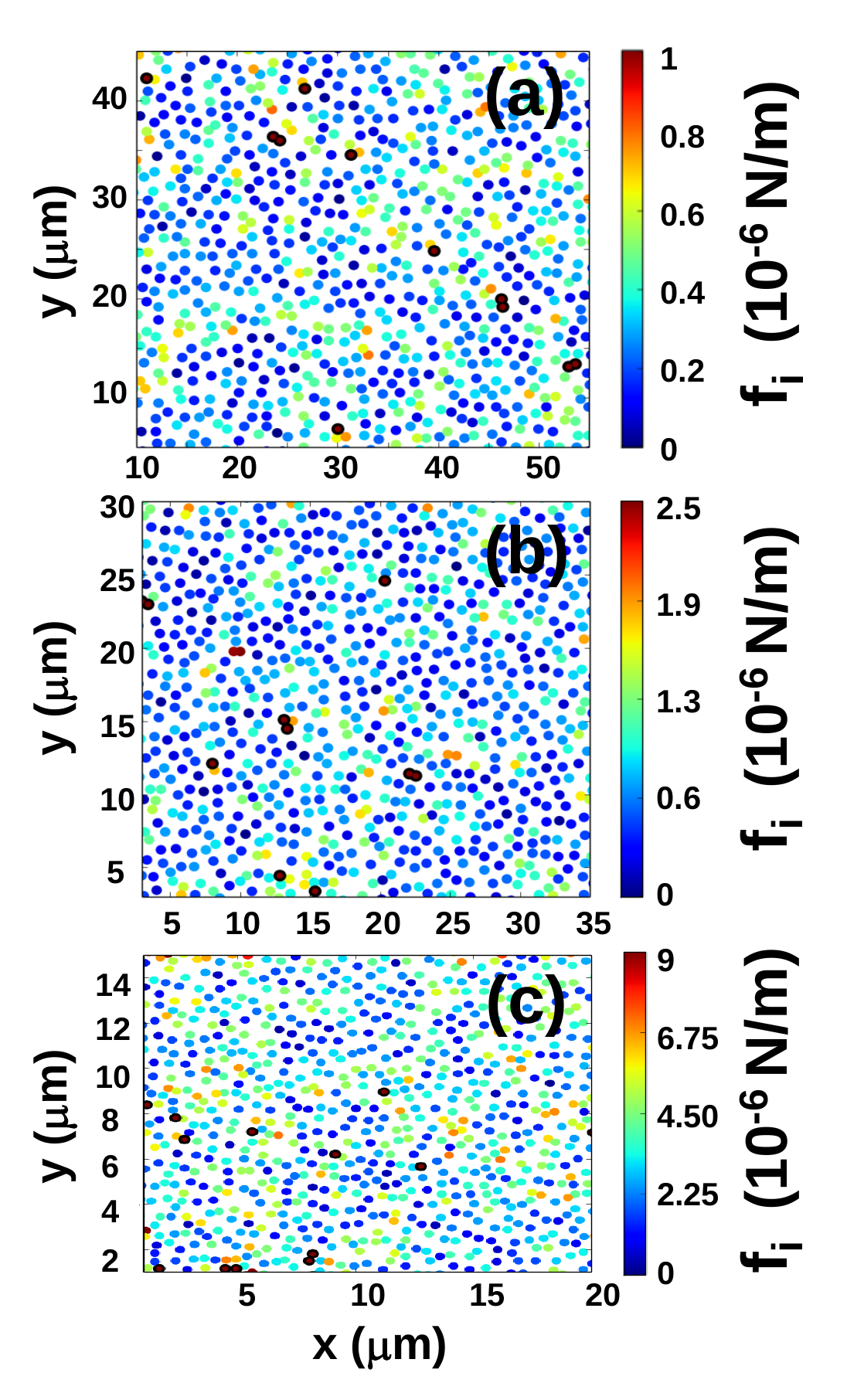}
\caption{ Color-coded maps for the modulus of the inter-vortex
repulsive force for vortex densities of (a) 9, (b) 18.8, and (c)
59.4\,G corresponding to the decoration images of
Fig.\,\ref{figure2}. Vortices with modulus of the inter-vortex
repulsive force larger than the mode value  plus  half-width at
half-maximum  are highlighted in black. \label{figure4}}
\end{figure}

In  spite of the absence of important fluctuations in vortex
density, the pinning potential generated by a dense distribution of
CD produces a strong impact in the lattice structural properties,
particularly evident in the spatially inhomogeneous inter-vortex
interaction energy  and vortex-defect force. We will focus our study
in the  last magnitude since from its probability distribution we
will get the experimental information considered as input to
evaluate the typical energy barriers considered in the freezing
dynamics model we propose in Section IV. The vortex-defect force is
related to the inter-vortex repulsive force, $\mathbf{f_{\rm i}}$.
Figure \ref{figure4} shows maps of the magnitude of this force per
unit length for each vortex $i$, $f_{\rm i} \equiv \mid
\mathbf{f_{\rm i}} \mid$, computed as the modulus of

\begin{equation}
\mathbf{f}_{\rm i} = \sum_{\rm j} \frac{2 \varepsilon_{\rm
0}}{\lambda_{\rm ab}} \frac{\mathbf{r}_{\rm ij}}{\mid
\mathbf{r}_{\rm ij}\mid}K_{1}(\frac{\mid r_{\rm ij}
\mid}{\lambda_{\rm ab}}). \label{eq2}
\end{equation}

\noindent $K_{1}$ is the first-order modified Bessel function,
$\lambda_{\rm ab}$ the in-plane penetration depth and
$\varepsilon_{\rm 0}=(\Phi_{0}/4\pi\lambda_{\rm ab})^{2}$ the vortex
line-tension. The sum is performed up to a cut-off radius $r_{\rm
cut}=10 a_{0}$ since $\mathbf{f}_{\rm i}$  does not change
significantly when including terms at
 larger distances.  For every magnetic field, $\mathbf{f}_{\rm i}$ is calculated
 considering the value of the
 penetration depth  at the temperature at which the
vortex structure is frozen that we approximate by $T_{\rm irr}$. We
have considered the $\lambda_{\rm ab}(T/T_{\rm c})$ evolution
reported in Ref.\,\onlinecite{Keeshabilitation} for pristine
Bi$_{2}$Sr$_{2}$CaCu$_{2}$O$_{8}$ samples and calculated
$\lambda_{\rm ab}(T_{\rm irr}/T_{\rm c})$ from the
 data of the vortex phase diagram shown in Fig.\,\ref{figure1};  $\lambda_{\rm ab}$ for samples
with a density of CD of $B_{\rm \Phi}=5$\,kG is within 1\% this
value.~\cite{vanderBeek2000}

\begin{figure}
\includegraphics[width=\columnwidth]{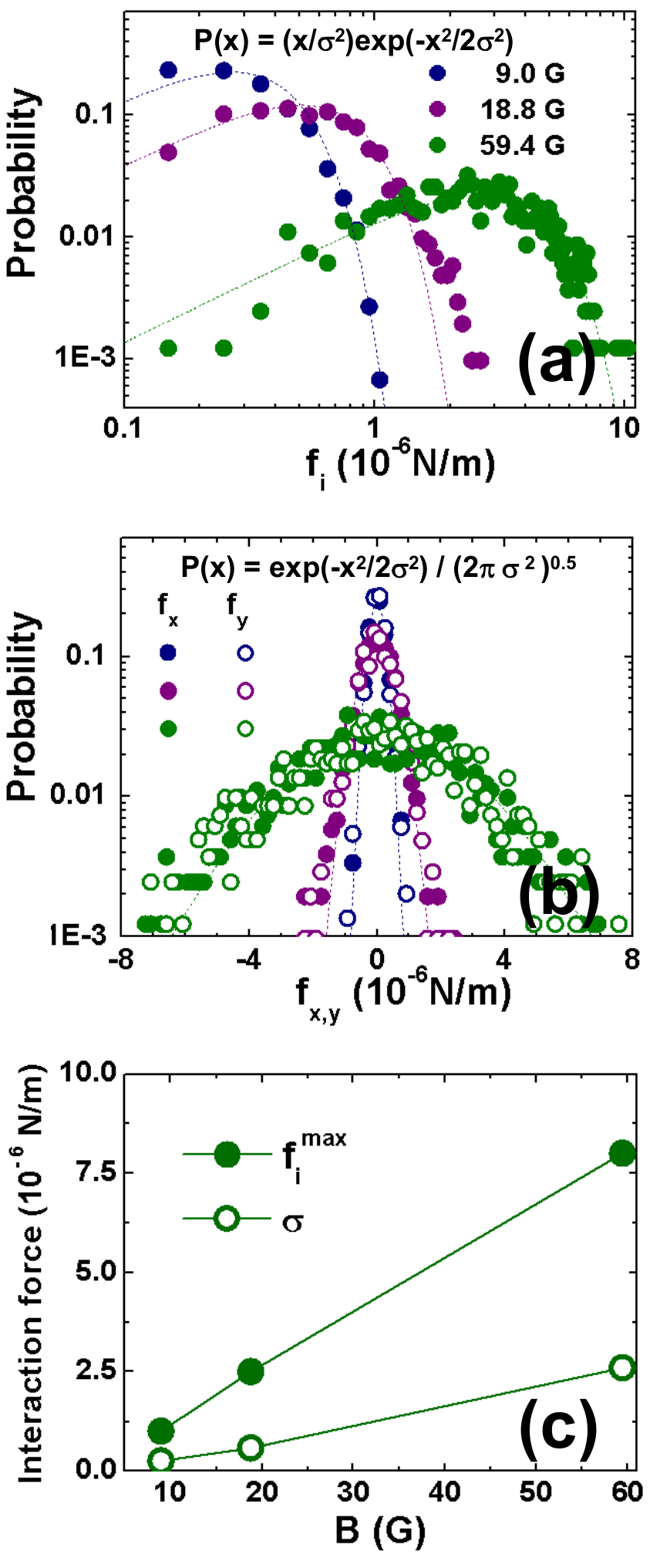}
\caption{Distributions of the inter-vortex repulsive force per unit
length for Bi$_{2}$Sr$_{2}$CaCu$_{2}$O$_{8}$ samples with a dense
distribution of CD at different vortex densities. (a) Modulus of the
inter-vortex force (points) and fits to the data with the Rayleigh
function indicated (dotted lines). (b) Distributions of the
$x$-(full points) and $y$-(open points) components of the
inter-vortex force and fits with a symmetric Gaussian distribution
(full lines). (c) Standard deviation and maximum values of the
inter-vortex force modulus distributions as a function of vortex
density.
 \label{figure5}}
\end{figure}

Since the vortex structures frozen during  field-cooling processes
are close to  metastable equilibrium  at $T_{\rm freez}(B)$,
non-zero values of $f_{\rm i}$ can only be ascribed to the force
exerted by the CD pinning potential to individual vortices.
Therefore the $ f_{\rm i}$ maps are an accurate estimation of the
minimum vortex-defect force at the local scale. These maps are
highly inhomogeneous and present clusters with larger inter-vortex
force magnitude. The $f_{\rm i}$ histograms of
Fig.\,\ref{figure5}\,(a) show that the mode value monotonically
enhances with increasing vortex density. In addition, the
distributions are not symmetric and have a larger weight in the
high-force part. On increasing $B$,  the $f_{\rm i}$ distributions
broaden significantly and get more asymmetric.

Figure\,\ref{figure5}\,(b) shows the histograms of the $x$ and $y$
components of the interaction force, $f_{\rm x}$ and $f_{\rm y}$. In
contrast to the $f_{\rm i}$ distributions, the force-components
distributions are symmetric with respect to zero and their width
increases with field. The $f_{\rm x}$ and $f_{\rm y}$ distributions
are properly fitted with Gaussian functions (see dotted lines),
implying that the individual components of the force varies at
random. Therefore, a Rayleigh functionality should be expected for
the distribution of the modulus of the force. The fits shown with
dotted lines in Fig.\,\ref{figure5} (a) (see the mathematical
expression on the top) indicate that the distributions of
$\mathbf{f_{\rm i}}$ follow reasonably well this functionality.
Therefore the increasing asymmetry of the $ f_{\rm i}$ distributions
with $B$ comes from the increment of the standard deviation of the
Gaussian distributions that follow the components of the force.

Finally, Fig.\,\ref{figure5}\,(c) shows that the dispersion obtained
from  the Rayleigh fits to the $f_{\rm i}$ distributions, $\sigma$,
as well as the  maximum value of the defect-vortex pinning force,
$f_{\rm i}^{\rm max}$, increase monotonically with field.
Considering that the CD pinning force is finite and has a maximum
value of $f_{\rm c}=U_{\rm 0}/r_{\rm r}$, extrapolating to higher
fields the data shown in Fig.\,\ref{figure5} (c) yields $f_{\rm
i}^{\rm max} \sim f_{\rm c}$ at around 4000\,G. This suggests a
crossover towards vortex configurations with ``instersticial
vortices'', i.e. not located on a CD, at fields $B \sim 0.8 B_{\rm
\Phi}$.

Assuming that localized vortices relax the local excess stress by
simple single-vortex excitations, linear extrapolation of $f_{\rm
i}^{\rm max}$ also allows the estimation of the field at which
half-loop excitations are exhausted in favor of double-kink
excitations~\cite{Nelson1993}. This occurs when $f_{\rm i}^{\rm max}
\sim f_d \approx U_0/d = f_c (d/r_{\rm r}) \approx f_c/20$. Hence,
for $B \sim 0.04 B_{\rm \Phi}$ we predict the extinction of
half-loops at very short times. Comparing with the maximum field we
analyze, $B=0.01 B_{\rm \Phi}$, we can argue that relaxation will be
dominated by double-kinks first and by superkinks only in the
long-time limit.  This kind of arguments, combined with the
magnitude $f_{\rm i}^{\rm max}(B)$ obtained from the vortex
structures frozen during field-cooling magnetic decoration
experiments allowed us to estimate the typical energy barriers and
relaxation times considered in the freezing dynamics model presented
in the next section.

\section{Freezing dynamics model}

In this section we discuss the experimental results in terms of a
simple model for the dynamics of freezing of the vortex structure
during  field-cooling in the presence of the strong disorder
associated to a dense distribution of CD. This model allows us to
estimate the lifetime of the observed metastable vortex
configurations and  $T_{\rm freez}$ from characteristic parameters
of the system and considering information obtained from magnetic
decoration images taken at low fields and 4.2\,K (see
Fig.\,\ref{figure2}). We will assume that columnar disorder is
strong and dense ($B \ll B_{\Phi}$) such that, for temperatures of
the order of $T_{\rm freez}$, well defined vortex lines remain most
of the time individually pinned in a single columnar defect. Namely,
in this model we neglect the effect of point disorder.

During the freezing process, pinned vortices form metastable
configurations $\alpha$ characterized by the set of occupation
numbers of each CD, $\alpha\equiv \{n_k\}$, where $n_k=1$ if the
$k$-th CD is occupied by a vortex, and $n_k=0$ otherwise. In each
metastable state the interaction force $f_{\rm i}$ must balance in
average the defect-vortex force on vortex $i$. During the lifetime
of the metastable state, controlled by thermal activation, vortices
are bound to their columnar defects, only performing small futile
fluctuations. Near freezing, the lifetime of metastable states are
expected to be comparable to experimental times, and their typical
configurations similar to the ones observed by magnetic decoration.
We are thus considering that cooling down to 4.2\,K for magnetic
decoration has only the effect of further stabilizing the metastable
configuration frozen at $T_{\rm freez}$.  We will use this criterion
for inferring information about the dynamics near the freezing of
the observed vortex structures and ultimately unveil the relevant
excitations that will allow us to discern on the nature of the
frozen vortex structure.

In order to model the non-stationary dynamics connecting different
metastable states after a temperature quench, we will assume that
the relaxation process is mainly dominated by single vortex hopping
between the randomly-distributed CD. We are thus neglecting multiple
vortex hopping, which might be important at higher vortex densities.
For simplicity, we will consider identical CD, the so-called
non-dispersive case.~\cite{Blatter} Under this assumption, the
contribution of super-kink excitations is not relevant and thus the
optimal thermal excitations of an individual vortex line are either
half-loops or double kinks (DK).~\cite{Blatter} These excitations
allow a vortex to escape from one columnar defect, and to be then
retrapped by a nearby defect located at a typical distance $d \sim
(\Phi_{\rm 0}/B_{\rm \Phi})^{1/2} \ll a_{\rm 0}$. Since we also have
$d \ll \lambda_{\rm ab}(0) < \lambda_{\rm ab}(T)$, the vortex-defect
force  on vortex $i$ is practically the same for consecutive
metastable states. If we denote the metastable state at a given time
by a supraindex $\alpha$, the relaxation process connecting
different metastable states can thus be effectively viewed as a
non-steady transport process driven by the heterogeneous set of
average local forces associated to the metastable state, $\{{\mathbf
f}^{\alpha}_{\rm i} \}$. The dependence of $f_{\rm i}$ with $\alpha$
allow us to  include the non-steady variation of vortex-vortex
interactions during the relaxation and its magnetic field
dependence.

\begin{figure}[ttt]
\begin{center}
\includegraphics[width=\columnwidth]{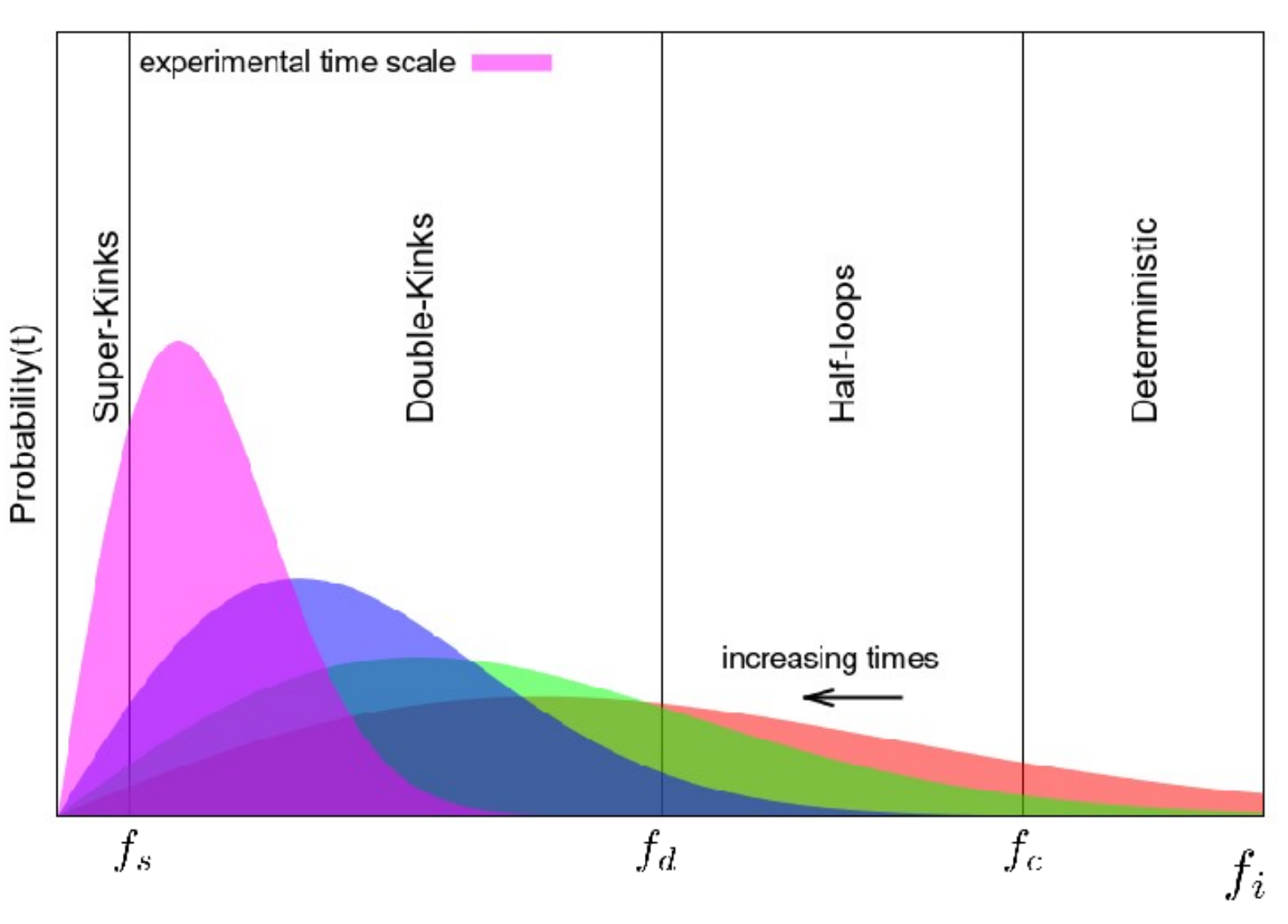}
\end{center}
\caption{Schematic picture for the distribution of the local force
relaxation process in the limit $d \ll a_{\rm 0}, \lambda_{\rm ab}$.
The system starts from a highly disordered state with a broad
distribution of local forces. As relaxation goes on, the force
distribution becomes narrower and the system gets quickly trapped
into metastable states when all forces lie below $f_{\rm c}$.
Thermally-activated optimal hops then allow vortices to escape from
their CD via different mechanisms: half-loops for $f_{\rm d}< f <
f_{\rm c}$, double-kinks for $f_{\rm s} < f < f_{\rm d}$, and
super-kinks for $f<f_{\rm s}$ (for identical CD $f_{\rm s}=0$),
where $f_{\rm c}$ and $f_{\rm d}$ depend on the pinning energy,
radius and separation between CD and $f_{\rm s}$ also depends on the
degree of CD disorder. During a field-cooling process, the
distribution becomes practically frozen at a temperature $T_{\rm
freez}$ for the experimental time scale.} \label{fig:schematic}
\end{figure}

Half-loops excitations are expected to be relevant for pinned
vortices such that $f_{\rm d} < f^{\alpha}_i < f_{\rm c}$, where
$f_{\rm d} \approx U_0/d$ is the force at which a half-loop involves
a displacement of the order of the average separation between
defects, $d$, and $f_{\rm c} \approx U_0/r_{\rm r}$ is the critical
depinning force from a single defect. As we show below, half-loops
drive the relaxation at very short time-scales; for larger times
most local forces $f^{\alpha}_{\rm i}$ drop below $f_{\rm d}$ and
half-loop excitations are exhausted. Figure\,\ref{fig:schematic}
shows a schematic representation of the relevant excitations at each
time-scale. Therefore, at large time-scales relaxation is mainly
driven by DK excitations since most of vortices feel an interaction
force $f_{\rm i} < f_{\rm d}$ from the other vortices. Optimal DK
have a longitudinal length $z \approx d\sqrt{\epsilon_{\rm l}/U_0}$
and cost an energy $2E_{\rm K} \approx 2d\sqrt{\epsilon_{\rm l}
U_0}$. The line tension $\epsilon_{\rm l} \sim
\varepsilon_0/\Gamma^2$ is normalized by the anisotropy of the
vortex system $\Gamma$, and corresponds to the limit of
short-wavelength distortions in the $c$ direction involved in the
kink formation. The proliferation of optimal DK excitations allows a
vortex localized at a defect to hop to a neighbor one at a distance
of the order of $d$. Provided that the energy barrier for such
excitation, $U_{\rm i}^{\rm \alpha}=(2E_{\rm K} - f_{\rm i}^{\rm
\alpha} d z)$, is larger than $k_{\rm B} T$, the typical time to
escape from the defect is given by the Arrhenius law $\tau_{\rm
i}^{\rm \alpha} = \tau_0 e^{U_{\rm i}/k_{\rm B} T}$, yielding

\begin{equation}
  \tau_{\rm i}^{\alpha} \approx
  \tau_0\;e^{(2d\sqrt{\epsilon_{\rm l} U_0} - d^2\sqrt{\epsilon_{\rm l}/U_0}f^{\alpha}_{\rm i})/k_{\rm B} T},
\label{eq:escapevortexi}
\end{equation}

\noindent where $\tau_0$ is a characteristic time, or inverse of the
attempt frequency. The last formula is an estimate for the
\textit{escape time} of a vortex $i$ feeling the interaction force
$f^{\alpha}_{\rm i}$   with all the other vortices \textit{on a
given metastable state} $\alpha$.

We can define the lifetime of a given metastable configuration
$\alpha$  as the   minimum single-vortex escape-time among all
vortices, as one vortex hop changes a pair of CD occupation numbers,
producing a new metastable state $\alpha'$. Such a minimal escape
time corresponds to the minimal escape barrier $U^{\rm
\alpha}=\min_{i}[2d\sqrt{\epsilon_{\rm l} U_{\rm 0}} -
d^2\sqrt{\epsilon_{\rm l}/U_{\rm 0}}f^{\rm \alpha}_{\rm i}]$.
Therefore, it corresponds to the escape time of the least bounded
vortex, feeling the maximum force $f^{\rm \alpha}_{\rm
max}=\max_i[f^{\rm \alpha}_i]$ in the metastable configuration
$\alpha$. Remembering that $U_{\rm 0} \sim \varepsilon_{\rm 0}
(r_{\rm r}^2/2\xi^2)$, we finally get the escape time for a given
metastable configuration

\begin{equation}
  \tau^{\rm \alpha} \approx \tau_{\rm 0}
  \exp\left[\frac{U^{\rm \alpha}(a_{\rm 0}, \lambda_{\rm ab}, d, r_{\rm r})}{k_{\rm B}
  T}\right],
\label{eq:escapetime}
\end{equation}

\noindent where we have defined the effective energy barrier
associated with a given pinned configuration $\alpha$,

\begin{equation}
 U^{\rm \alpha} \equiv U^{\rm \alpha}(a_{\rm 0}, \lambda_{\rm ab}, d, r_{\rm r}) = \left(1  - \frac{{\tilde f}^{\rm \alpha}_{max}(a_{\rm 0}, \lambda_{\rm ab}) d \lambda_{\rm ab}}{2 \kappa^2 r_{\rm r}^2}  \right) 2E_{\rm
 K},
\label{eq:escapebarrier}
\end{equation}

\noindent as a function of the characteristic lengths $a_{\rm 0}$,
$\lambda_{\rm ab}$, $d$ and $r_{\rm r}$. The dimensionless force
${\tilde f}^{\rm \alpha}_{\rm max} = f^{\rm \alpha}_{\rm max}
\lambda_{\rm ab}/2\varepsilon_{\rm 0}$, is defined as

\begin{equation}
{\tilde f}^{\rm \alpha}_{\rm max}(a_{\rm 0},\lambda_{\rm ab}) =
\max_{i} \left|\sum_{j\neq i} K_1({\bf r}_{\rm ij}/\lambda_{\rm
ab}) \frac{{\bf r}_{\rm ij}}{r_{\rm ij}}\right|,
\label{eq:adimensionalforce}
\end{equation}

\noindent  obtained from the $\mathbf{f_{\rm i}^{\rm max}}$ data
shown in Fig.\,\ref{figure5} (c). Therefore $\alpha$ corresponds to
the metastable state frozen during the field-cooling decoration
experiment. The DK energy-cost can be written in terms of the
characteristic lengths $\lambda_{\rm ab}$, $d$, and $r_{\rm r}$ as

\begin{equation}
  2 E_{\rm K} \equiv 2 E_{\rm K}(\lambda_{\rm ab}, d, r_{\rm r}, \Gamma, \kappa) = \sqrt{2} \frac{\kappa r_{\rm r} d\varepsilon_{\rm 0} }{\lambda_{\rm ab} \Gamma}.
\label{eq:kinkcost}
\end{equation}

\begin{figure}[ttt]
\begin{center}
\includegraphics[width=\columnwidth]{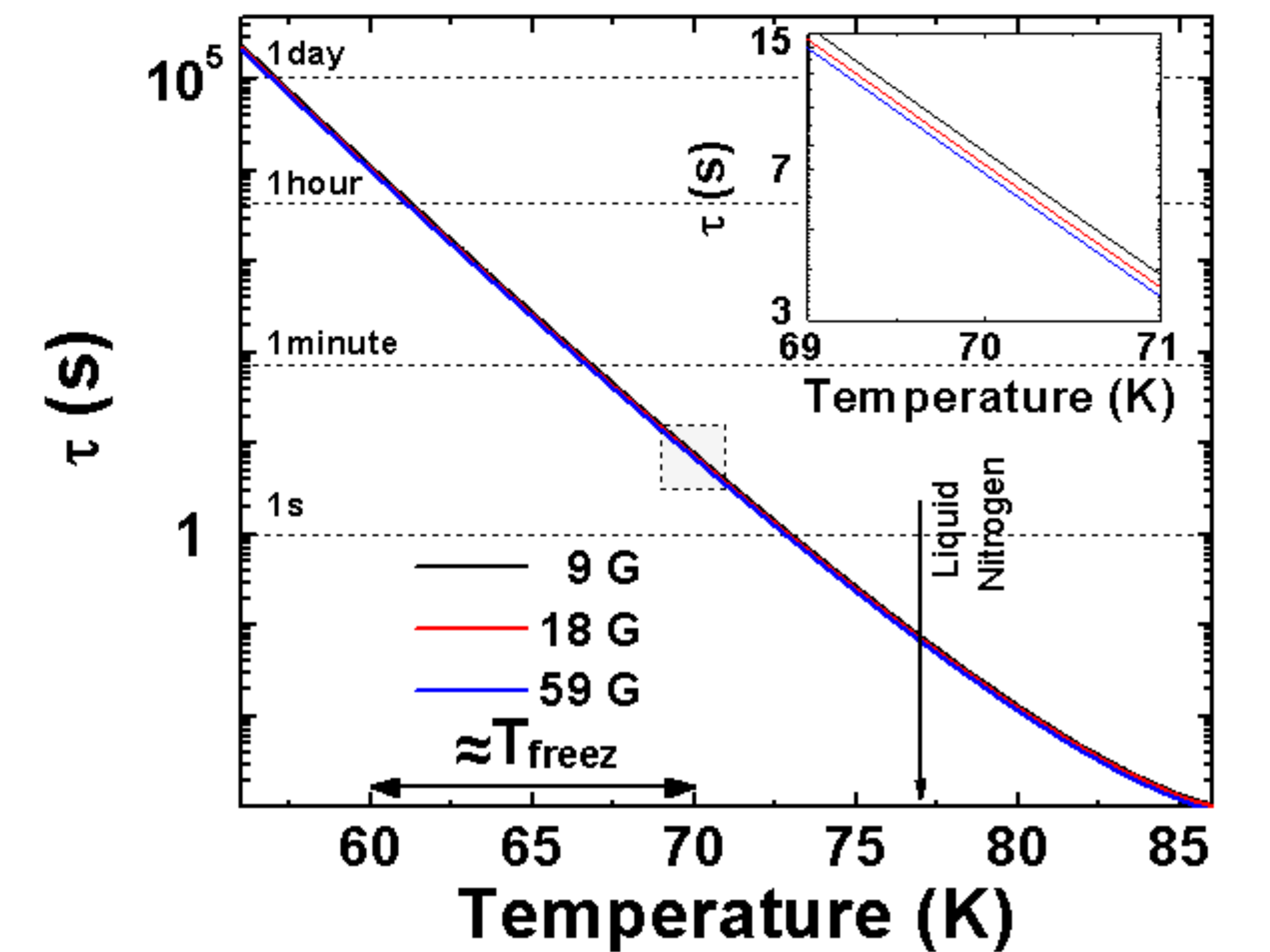}
\end{center}
\caption{Lifetime of the decorated metastable vortex configuration
vs. temperature, assuming only double-kink excitations during the
out of equilibrium relaxation. We indicate the estimated freezing
temperature $T_{\rm freez}$ at which the vortex structure is frozen
at lengthscales of $a_{\rm 0}$ as revealed by magnetic decoration.
The inset shows a weak decrease of the lifetime with increasing the
magnetic field for the field-range analyzed. } \label{fig:tauvsT}
\end{figure}

\noindent Within this model, the field-dependence of the lifetime
 comes from the field-dependence of the maximum value of the
defect-vortex force distribution for the given frozen configuration.
The insert to Fig.\,\ref{fig:tauvsT} shows that the lifetime weakly
decrease with increasing the magnetic field. Extrapolation of the
behaviour of Fig.\ref{figure5}(c) to larger fields suggests that
this trend continues, and that $T_{\rm freez}$ should therefore also
decrease with increasing magnetic field, following qualitatively the
behaviour of $T_{\rm {irr}}$ in Fig.\ref{figure1}.

\begin{table}[ttt]
\label{TableI}
\begin{tabular}{|c|c|c|c|c|c|}
\hline $B$ {[}G{]} & $\lambda_{\rm ab}(T_{\rm irr}) [\mu m]$ &
$T_{\rm irr} [K]$ & ${\tilde f}_{\rm max}(\lambda_{\rm ab}(T_{\rm
irr})) $ & $U/k_B[K]$ & $\tau[s]$ \\
\hline 9           & 0.69 &  85.3   & 0.6   & 26   & 0.0013\\
\hline 18.8        & 0.59 &  83.9   & 0.97  & 41   & 0.0016 \\
\hline 59.4        & 0.39 &  79.6   & 0.87  & 142  & 0.0059 \\ \hline
\end{tabular}
\caption{ Estimation of the lifetime $\tau$ of the magnetically
decorated metastable vortex configurations frozen during
field-cooling at $ \sim T_{\rm irr}$ and assuming only DK
excitations. We have used the following  sample characteristics:
 $d=0.065\;10^{-4}cm$, $r_{\rm r}=3.5\; 10^{-7}cm$,
($d/r_{\rm r}\sim 20$), $\lambda(T=0)=0.18\;10^{-4}cm$, $T_{\rm
c}=86.7$\,K, $\kappa \approx 200$, $\Gamma \approx 150$. The
parameters $\lambda_{\rm irr}=\lambda(T_{\rm irr})$ and
$\epsilon^{irr}_0=(\Phi_0/4\pi \lambda_{\rm irr})^2$. The
characteristic value $\tau_0 \sim 10^{-3}s$ is taken from the
frequency-dependent ac creep measured in samples from the same batch
than ours.~\cite{vanderbeek1995}} \label{table:tablairr}
\end{table}

Let us first start estimating the lifetime $\tau$  of the decorated
metastable vortex configurations ---we will drop the supraindex
$\alpha$ when referring to the frozen metastable state--- by making
again the reasonable assumption that they were frozen at, or very
near to, the irreversibility line, namely $T_{\rm freez}\approx
T_{\rm irr}$. From expressions of equations (\ref{eq:escapetime}),
(\ref{eq:escapebarrier}), (\ref{eq:adimensionalforce}), and
(\ref{eq:kinkcost}), we get the results for the  escape time $\tau$
shown in Table \ref{table:tablairr}, using as input the
dimensionless force ${\tilde f}^{\rm \alpha}_{\rm max}$ evaluated
from the experimental data and the sample characteristics indicated
in the caption of the table. In all cases we verified that
$\tilde{f}_d(\lambda_{\rm irr}) > {\tilde f}_{\rm max}(\lambda_{\rm
irr})$ so that half-loop excitations are not relevant. The estimated
lifetimes of Table \ref{table:tablairr}, of the order of $\tau_{\rm
0}\sim 10^{-3}$\,s, are much smaller than the typical time-scales of
decoration experiments, namely $\tau(T_{\rm irr})\ll \tau_{\rm
exp}$, with $\tau_{\rm exp}$ of the order of seconds or minutes.
Therefore, we should necessarily have $T_{\rm freez}<T_{\rm irr}$
since the freezing temperature should correspond to $\tau(T_{\rm
freez})\sim \tau_{\rm exp}$. Forcing such a condition,  from
equations (\ref{eq:escapetime}), (\ref{eq:escapebarrier}),
(\ref{eq:adimensionalforce}), and (\ref{eq:kinkcost}) we can
estimate the value of $T_{\rm freez}$. In order to do this
realistically, we need to know the temperature-dependence of
$\lambda_{\rm ab}$ at $T<T_{\rm irr}$. Here, we will simply assume
the dependence $\lambda_{\rm ab}(T) = \lambda_{\rm
ab}(0)/\sqrt{1-(T/T_{\rm c})^4}$ with $\lambda_{\rm ab}(0)=180$\,nm.
This dependence yields a  reasonable analytical approximation for
the values reported at $T \sim T_{\rm irr}$~ \cite{Keeshabilitation}
and shown in Table~\ref{table:tablairr} for $T=T_{\rm irr}(B)$ for
the $B$ studied here.

 Recalculating
${\tilde f}_{\rm max}^{\rm \alpha}$ and $U^{\rm \alpha}$ as a
function of $T$ using this approximation for $\lambda_{\rm ab}(T)$,
and setting $\tau(T_{\rm freez})\sim \tau_{\rm exp}$, with
$\tau_{\rm exp}$ ranging from seconds to hours, we get freezing
temperatures in the range $T_{\rm freez}\approx 60-70$\,K as shown
in Fig. \ref{fig:tauvsT}.  The insert to this figure shows that the
field-dependence of the lifetime, and thus of $\tau(T_{\rm freez})$,
is very weak. Indeed the experimentally-detected change in almost
one order of magnitude in $\mathbf{f_{\rm i}^{\rm max}}$ on
increasing field only impacts in changing $\sim 5$\,\% the escape
time at $T_{\rm {irr}}$, a magnitude that is even smaller at lower
temperatures due to the decrease of $\lambda_{\rm ab}(T)$.
 It is also interesting to note that at the liquid
nitrogen temperature $T_{\rm N}$, at which the sample spends some
minutes during the magnetic decoration cooling protocol down to
4.2\,K, the dynamics is relatively fast, with lifetimes of roughly
milliseconds. This means that the frozen configuration  has a memory
of a temperature $T_{\rm freez} < T_{\rm N}$, indicated in
Fig.\,\ref{fig:tauvsT}.

We conclude our theoretical analysis discussing the validity of some
of the main assumptions of our model. We have assumed that CD are
perfectly parallel to the applied magnetic field, so vortices
individually localize in a single defect before a thermally
activated event drives it to a neighbor, more favorable pin.
However, in the experiments there might exist a misalignment between
the applied field and the CD direction of less than 5 degrees. At
low temperatures, below the irreversibility line, pinned vortices
aligned with a single CD are stable only below the characteristic
tilting ``locking angle'' $\theta_L$, between the applied field and
the defects. For misalignements larger than $\theta_L$ vortices can
still feel the effect of CD but develop a kinked structure
(staircase like) connecting more than one defect.  Only above a
larger characteristic ``trapping angle'' $\theta_t$, vortices are
not locked by CD and the response to a tilt becomes
linear.~\cite{Nelson1993} For Bi$_{2}$Sr$_{2}$CaCu$_{2}$O$_{8}$
irradiated with a $B_{\rm \Phi}= 5$\,kG dose, for $T \sim T_{\rm
irr}$ and $B \ll B_{\rm \Phi}$, our experimental conditions, a
trapping angle $\theta_t \approx 60-75$ degrees was experimentally
determined in Refs.~\onlinecite{vanderbeek1995,Seow1996}. Being CD
efficient to pin and deform the vortices even for such large angles,
their effect for misalignments one order of magnitude smaller should
be much stronger. Up to our knowledge the locking angle $\theta_L$
(below which perfect alignment is expected as it is assumed in our
model), has not been experimentally determined for the irradiation
dose of our sample. However, it can be theoretically estimated  as
$\theta_L = (4 \pi \epsilon_l/\Phi_0 H) \theta_t$.~\cite{Blatter}
Using the reported experimental value $\theta_t \approx 60-75$
degrees, and the line tension $\epsilon_l$ corresponding to our
material, we get $\theta_L \approx 3-15$ degrees for the three
fields we have studied. It is worth noting that these values are in
accordance with the experimental value $\theta_L \approx 20$ degrees
measured for YBa$_2$Cu$_3$O$_7$ with the same dose, at fields and
temperature similar to ours \cite{Silhanek2003}, if we take into
account explicitly the anisotropy ratio of both systems. Since
$\theta_L \sim 3-15$ degrees is of the order or larger than our
experimental uncertainty of $5$ degrees, we conclude that the
perfect alignment assumption of our model is fairly acceptable.
Indeed, even for tilting angles $\theta \gtrsim \theta_L$ the
typical distance between these kinks is expected to diverge as
$(\theta - \theta_L)^{-1/2}$ as we approach
$\theta_L$.~\cite{Nelson1993} Therefore, perfect alignment is then
possible even for $\theta \gtrsim \theta_L$ due to a finite size
effect, when the distance between kinks become of the order of the
sample width, in our case $\sim 50\,\mu$m.

In our model we have neglected the contribution of half-loop
excitations. As argued in the previous section this is justified
from the fact that these excitations are important for forces
smaller than ${\tilde f}_c = (U_0/r_{\rm
r})(\lambda_{ab}/2\epsilon_0) \approx 100$ and larger than ${\tilde
f}_d = {\tilde f}_c (r_{\rm r}/d) \approx {\tilde f}_c/20 \approx 5$
at $T_{\rm {irr}}$, which is larger than the adimensional forces of
Table \ref{table:tablairr} for the range of magnetic fields
analyzed. Half-loops would thus drive the relaxation only at very
short times after the quench. We have also neglected the dispersion
in the CD pinning energy. This dispersion is produced both by the
on-site dispersion, (typically arising from the unavoidable
dispersion in the ions tracks diameters), or by the inter-site
dispersion (arising from the dispersion in the nearest neighbor
distance between tracks, which is indeed expected to be
Poisson-distributed). According to Ref.\,\onlinecite{Soret2000}, the
main source of energy dispersion comes from the diameter dispersion,
which for 5.8 GeV Pb ions was estimated to be of the order of
$15\%$. This implies an energy dispersion $\Delta U_0/U_0 \sim
2.5\%$ at zero temperature. The crossover force $f_s$ from
double-kink to superkink excitations is thus of the order of $f_s
\approx f_d (\gamma/U_0) \approx f_d (\Delta U_0/U_0) = f_d/40$.
Since for our case $f_d = U_0/d \approx 20 2\epsilon_0/\lambda $, we
get the adimensional force ${\tilde f}_s \equiv f_s
(\lambda_{ab}/2\epsilon_0)  \sim 0.5$. From these estimates we first
note that the characteristic force extracted from magnetic
decoration images  is ${\tilde f}_{max} \sim 1$ (see table
\ref{table:tablairr}). Since ${\tilde f}_s \sim {\tilde f}$, the
typical hopping distance in the variable-range-hopping on
netoregime~\cite{Blatter} becomes $u_{\bm VRH} \approx d
(f_v/f)^{1/3} \sim d$. We thus conclude that the dominance of DK
excitations between columnar tracks separated by a typical distance
$d$ in the direction of the local force is a fair approximation for
the most common elementary excitations driving the relaxation near
$T_{\rm freez}$.

Finally, we have also neglected the possible thermal renormalization
of the pinning potential expected to become important above a
characteristic temperature~\cite{Nelson1993} $T_1 \approx T_{\rm c}
(r_{\rm r}/4\xi) \sqrt{\ln \kappa/Gi}/(1+(r_{\rm r}/4\xi) \sqrt{\ln
\kappa/Gi})$. Using parameters for
Bi$_{2}$Sr$_{2}$CaCu$_{2}$O$_{8}$, $Gi=10^{-2}$, $\kappa=200$,
$r_{\rm r}=35$ nm and $\xi=1$\,nm we get $T_1 \approx 72$\,K, which
is of the order of the $T_{\rm irr}$ for the range of magnetic
fields analyzed but larger than our estimated $T_{\rm freez}$.
Therefore, thermal renormalization of the pinning energies is
expected to be weak, justifying our approximation.

\section{Discussion}

The predictions from the model discussed in the previous section,
based on quantitative information obtained from decoration
experiments at 4.2\,K, suggest that the decorated structures were
dynamically frozen at a much larger temperature $T_{\rm freez}$ by
the strong decrease of vortex mobility with temperature induced by
the CD dense pinning potential. Within the model, this mobility is
dominated by Arrhenius activation through the temperature-dependent
\textit{finite} barriers associated with DK excitations, as barriers
for half-loops can be overcame in typical times of the order of
milliseconds while the larger barriers expected for super-kinks were
disregarded on the basis that DK alone are able to yield metastable
states with macroscopic lifetimes. Within this scenario, it is worth
mentioning that previous works studying samples with $B_{\rm
\Phi}=5$\,kG from the same batch than ours reported that the
low-field regime presented non-divergent activation barriers varying
linearly with the ac driving current.~\cite{vanderbeek1995}  This is
in contrast to the divergent barriers found at higher fields,
consistent with half-loop excitations. On the other hand, it was
proposed that for a high density of CD the change in pinning
energies when increasing field are negligible compared to the change
in interaction energy, giving place to a ``discrete superconductor''
picture.~\cite{vanderbeek2001} These two findings are consistent
with our model assumption that the relevant excitations driving the
non-steady relaxation are just DK, and that the role of super-kink
excitations is not important for the time-scales typical of our
magnetic decoration experiments.

Note that this is in contrast with the barriers expected for
 instance for the Bose glass phase, which
tend to diverge near equilibrium, a signature of the strong
localization of vortex lines at very long times with a broad
variable range hopping. The amorphous order observed in the
decorations can then be interpreted as a snapshot of a non-entangled
liquid-like structure metastable at $T_{\rm freez}$, a temperature
located below the first-order transition line. This non-entangled
highly viscous and {\it far from equilibrium} liquid-like structure
contrasts with the quasi-long-range positionally-ordered vortex
structures observed in pristine samples of the same compound at
$T<T_{\rm FOT}$.~\cite{Cejas2015}  Experiments with significantly
larger cooling times would allow to ascertain whether a more ordered
vortex structure is stable for a very dense distribution of CD.

Figure \,\ref{fig:schematic} shows a schematic picture for the
relaxation dynamics of the defect-vortex forces. Within our model,
the dynamics at $T_{\rm freez}$ is mainly controlled by DK
excitations since most of the defect-vortex forces satisfy $f_{\rm
s} < f < f_{\rm d}$, with $f_{\rm s} \sim \gamma/d$ the force below
which super-kink excitations dominates. The parameter $\gamma$ is
the dispersion in pinning energies coming from the differences
between columnar defects and from the disorder in their spatial
distribution. In our model we have assumed that $\gamma$ is
negligible compared to $U_{\rm 0}$, so $f_{\rm s}$ is very small and
super-kink dynamics would become relevant only at very large
time-scales. If $\gamma$ is not small, then the non-steady
relaxation may be dominated by variable range hops with DK becoming
inefficient to irreversibly drive the vortices to a lower energy
state.

A previous work~\cite{vanderbeek2001} proposes  that the low-field
vortex state nucleated in the case of a sample with a dense
distribution of CD would not differ fundamentally from that observed
in pristine samples. This is based in the finding that the energy
difference between two metastable states in the former case is
dominated by the vortex-vortex interaction energy rather than by the
differences in pinning energy. Within this view, each vortex line is
however confined and pinned to a CD. This proposal is similar to the
one stating that the conventional Bose glass phase may have a
crossover to a putative ``Bragg-Bose glass phase''.~\cite{Giamarchi}
The later is a glassy phase with quasi-long-range order but confined
in CD, and thus
 individual vortex lines are macroscopically flat, in sharp contrast
to the rough vortex lines of the Bragg-glass phase expected for weak
point-disorder. Interestingly, our analysis suggests that the truly
glassy relaxation dynamics in our samples, such as the super-kink or
variable range hopping dynamics, or collective creep dynamics made
of correlated hops (corresponding to one of the possible equilibrium
phases), would become dominant only in the limit of very long
relaxation times, {\it much larger} than the one probed during the
magnetic decoration quenching process. These experiments, as they
are currently done, can not give any information about the subjacent
equilibrium phase expected at very long-times. This will critically
depend on the degree of dispersion of the pinning energies, $\gamma
\sim f_{\rm s} d$.~\cite{Blatter}

Therefore, it would be interesting to implement glass-annealing
techniques in order to reach configurations with a narrower
distribution of vortex-defect forces and thus with less memory of
the liquid phase and less accumulated stress. Observing the changes
in the translational order of the decorated lattice as a function of
the quenching time, for instance, may tell us which phase is more
plausible, as the corresponding equilibrium correlation length
slowly grows with time. In particular, if a topologically ordered
equilibrium phase exists at low fields in the presence of a high
density of CD or a ``discrete superconductor'' picture is valid,
then the density of dislocations in the vortex structure should
display a decrease with time. This effect might be seen, for
instance, by comparing the number of $\rho_{\rm def}$ detected in
magnetic decoration experiments performed at significantly different
cooling rates.

\section{Conclusions}

We have analyzed, through magnetic decoration images, dynamically
frozen vortex configurations in heavy-ion irradiated
Bi$_{2}$Sr$_{2}$CaCu$_{2}$O$_{8}$ samples with a very dense
distribution of columnar defects. For low vortex densities compared
with the CD density, we find an amorphous phase with liquid-like
correlations, and an approximately Gaussian defect-vortex force
distribution indicating a randomly-oriented pinning scenario. By
assuming vortices individually trapped at identical CD, we show that
the observed translational order is fairly consistent with a
relaxation dynamics dominated by DK excitations near the freezing
temperature $T_{\rm freez}$. Using a simple model  and input from
the $\mathbf{f_{\rm i}^{\rm max}}$ experimental data obtained from
magnetic decoration images, we predict a freezing temperature of the
same order but smaller than the irreversibility temperature. We
argue that magnetically-decorated structures hence correspond to a
typical configuration of a non-entangled vortex-liquid state with
strongly reduced mobility, rather than to a metastable state of an
equilibrium glassy phase with divergent relaxation barriers
associated to the localization of vortices in CD even at very long
times. Experiments with significantly larger cooling times or
glass-annealing techniques are mandatory in order to dilucidate if,
as na\"{i}vely expected,  the equilibrium vortex phase nucleated in
a very dense distribution of strong pins is a more ordered one.

\section{Acknowledgments}

This work was made possible thanks to the support of the
ECOS-Sud-MinCyT France-Argentina bilateral program, Grant No.
A09E03. Work done at Bariloche was partially founded by PICT-PRH
2008-294 and University of Cuyo Reserch Grant No. 06-C381. A.B.K,
D.D. and Y.F. also acknowledge support from ANPCyT-PICT-2011-1537.

\end{document}